\documentclass[twocolumn,amsmath,amssymb,nofootinbib,floatfix
]{revtex4}

\usepackage{amsmath,amssymb}
\usepackage{graphicx}
\usepackage{dcolumn}
\usepackage{bm}
\usepackage[colorlinks=true,linkcolor=blue,urlcolor=blue,filecolor=black,
citecolor=red,pdfstartview=FitV,pdftitle={},pdfsubject={},pdfkeywords={},
pdfpagemode=None,bookmarksopen=true]{hyperref}
\usepackage{graphicx}

\begin{document}

\preprint{APS/123-QED}

\title{Extraction of energy from a black hole in Einstein-Maxwell-scalar theory}

\author{Cheng-Yong Zhang\textsuperscript{1}}
$\email{zhangcy@email.jnu.edu.cn}$
\affiliation{
 Department of Physics and Siyuan Laboratory, Jinan University, Guangzhou 510632, China
}

\author{Zehong Zhang\textsuperscript{1}}
\email{zhangzh@stu2022.jnu.edu.cn, corresponding author}
\affiliation{
 Department of Physics and Siyuan Laboratory, Jinan University, Guangzhou 510632, China
}

\author{Ruifeng Zheng\textsuperscript{1}}
$\email{zrf2022@stu2022.jnu.edu.cn}$

\affiliation{
 Department of Physics and Siyuan Laboratory, Jinan University, Guangzhou 510632, China
}%

\begin{abstract}
Recently, it has been discovered that the nonlinear self-interaction of matter can induce energy extraction from black holes beyond superradiant instability. This process is closely associated with the occurrence of a dynamical first-order transition between different types of static black holes. To explore whether first-order phase transitions invariably lead to energy extraction, we have investigated the evolution of black holes in the Einstein-Maxwell-scalar model with a higher-order coupling. In this model, there are also dynamical first-order phase transitions between black hole solutions. Our findings indicate that energy can only be extracted from a small, stable hairy black hole in this model. However, this energy extraction is more closely related to the growth of the black hole horizon radius, rather than the dynamical transition between different types of black holes. This suggests that a dynamical first-order phase transition does not necessarily result in energy extraction.\\ \par


\end{abstract}

\maketitle


\section{\label{sec:level1}INTRODUCTION}

Black holes (BH) in asymptotically flat space are often thought of as completely dead classically, which means they only absorb radiation and energy, 
without radiating energy externally. 
However, in 1969, Penrose et al.\cite{Penrose:1969pc,Penrose:1971uk} proposed an interesting process that could extract energy from a rotating Kerr black  hole . The most essential reason is the existence
of ergoregions, where timelike particles can have negative energies.
Misner et al.\cite{Misner:1972kx,Futterman:1988ni}pointed out that waves can also extract rotational energy through superradiance in which an impinging wave is amplified as it scatters off a rotating black hole .  Superradiance requires dissipation that can be provided by the ergoregion \cite{Brito:2015oca}.    Many investigations of superradiance were concentrated on rotating black holes \cite{Brito:2015oca,Press:1972zz, Teukolsky:1974yv,Dolan:2008kf,Zhang:2014kna,Vicente:2018mxl,Benone:2019all}. But the discussions of superradiance have been extended to charged black holes with charged field perturbations \cite{Bekenstein:1973mi,Hod:2012wmy,Hod:2013nn,Zhang:2015jda,Zhu:2014sya,Konoplya:2014lha,DiMenza:2014vpa,Liu:2020evp,Zhang:2020sjh}, since a spacetime containing a
nonrotating charged black hole constitutes an effectively dissipative environment for charged fields.  Fully nonlinear investigations into superradiance, whether concerning charged or rotating black holes, are notably scarce. However, there exist significant exceptions that stand out in this field \cite{East:2013mfa, Baake:2016oku}, in which
superradiance is shown to occur at a full nonlinear level. 
As the incident wave energy increases, the amplification of the scattered waves, as well as the energy extraction efficiency from the black hole, is reduced.

Superradiance in black hole spacetimes can trigger instabilities.
Surrounding the rotating black hole with a reflecting
mirror, in this case, the wave would bounce back and
forth, between the mirror and the black hole, amplifying
itself each time, and giving rise to ``black hole bombs'' \cite{Press:1972zz, Cardoso:2004hs, Witek:2010qc,
Cardoso:2004nk,
Dolan:2012yt}.
This is exactly the same principle behind the instability of Kerr
black holes against massive bosonic perturbations because in this case
the mass of the field works as a barrier at infinity \cite{Damour:1976kh,Zouros:1979iw,Detweiler:1980uk}. 
The development of black hole bombs is studied nonlinearly around rotating black holes \cite{Okawa:2014nda,East:2017ovw,East:2018glu,Chesler:2018txn} and charged black holes \cite{Sanchis-Gual:2015lje,Sanchis-Gual:2016tcm,Bosch:2016vcp,Garcia-Saenz:2025rbc}.
With the backreaction of amplified waves on spacetime,
the black hole bomb induced by superradiant instability will inevitably terminate at some point and the unstable seed black hole must evolve into a stable state, which is usually a hairy black hole \cite{Hod:2012px,Dias:2011tj,Herdeiro:2014goa,Herdeiro:2016tmi}.


In addition to superradiant instability, by incorporating non-linear self-interaction into the massive charged complex scalar field in general relativity, 
two novel dynamical mechanisms were discovered recently, which can also trigger black hole bomb phenomena \cite{Zhang:2023qtn}. 
The nonlinear effect
of the scalar field can destroy the stability of the Reissner-Nordstr\"om (RN) black
hole and release substantial charge and energy to develop scalar hair.
The RN black hole undergoes a first-order phase transition and eventually evolves into a hairy black hole.
There is energy extraction and a first-order phase transition phenomenon in the model. 
Is this energy extraction behavior determined by a first-order phase transition? Or is the dynamical first-order phase transition always accompanied by energy extraction? To make this clear, one should consider other models with first-order phase transition, such as scalar-GB (sGB)  \cite{Liu:2022fxy} and Einstein-Maxwell scalar (EMS) models \cite{Zhang:2021nnn, Jiang:2023yyn, Zhang:2022cmu}.

Our   consideration here is the EMS model in an asymptotic flat spacetime. 
There are three branches of static black hole solutions in the EMS model: a branch for the RN black hole, a branch for the unstable scalarized black hole (SBH) and a branch for stable SBH \cite{Blazquez-Salcedo:2020nhs,LuisBlazquez-Salcedo:2020rqp}.  
RN black holes and stable SBHs have been shown to transform to each other through dynamical first-order phase transition, and unstable SBHs play the role of critical solutions at the transition threshold \cite{Zhang:2021nnn, Jiang:2023yyn, Zhang:2022cmu}.

The specific action we consider is
\begin{equation}
S=\frac{1}{16\pi } \int \mathrm{d}^{4} x\sqrt{-g} \Big[R-2\nabla  _{\mu }\phi \nabla ^{\mu } \phi -f(\phi )F_{\mu \nu } F^{\mu \nu } \Big]~,
\end{equation}
where $R$ is Ricci scalar, $F_{\mu \nu }$ represents the field strength of Maxwell field $A_{\mu }$, and the real scalar field $\phi$ nonminimally couples to the Maxwell invariant through the coupling function $f(\phi)=e^{\beta \phi ^{n} }$, 
in which $\beta$ is the  parameter that quantifies the strength of the coupling. This model offers a theoretical framework for exploring the influence of new degrees of freedom on the behavior of gravitational and electromagnetic fields. When $n=1$,  this model is commonly known as Einstein-Maxwell-dilaton theory, which is relevant to low-energy string theories, supergravity models, and holographic models \cite{Garfinkle:1990qj,Swingle:2017zcd,Sahoo:2023czj}. In recent years, the model with $n=2$ has attracted much attention due to the phenomenon of spontaneous scalarization, which has the potential to serve as a probe for testing the strong-field regime of gravity \cite{Herdeiro:2018wub,
Fernandes:2019rez,Astefanesei:2019pfq,Yu:2020rqi,Guo:2021zed,
Zhang:2021etr,
Luo:2022roz,
Xiong:2022ozw,
Guo:2023mda,Doneva:2022ewd}. More recently, there has been interest in studying models with higher order coupling functions, in which critical behaviors in the dynamical first-order transition between different kinds of static black holes were discovered \cite{Blazquez-Salcedo:2020nhs,LuisBlazquez-Salcedo:2020rqp,Zhang:2021nnn,Zhang:2022cmu,Jiang:2023yyn}. In this paper, we choose $n=4$ to explore the energy extraction process during the dynamical first-order phase transition.

The Einstein equations are given by 
\begin{equation}
R_{\mu \nu } -\frac{1}{2} Rg_{\mu \nu } =2\Big(T_{\mu \nu }^{\phi }+f(\phi )T_{\mu \nu }^{A }  \Big)~,
\end{equation}
where the energy-momentum tensor of the scalar and Maxwell fields are represented as:
\begin{equation}
\begin{aligned}
    T_{\mu \nu }^{\phi }&=\partial _{\mu } \phi \partial _{\nu } \phi -\tfrac{1}{2} g_{\mu \nu } \nabla _{\rho }\phi \nabla ^{\rho }\phi ,
\\
T_{\mu \nu }^{A }&=F _{\mu \rho }  F _{\nu }^{\rho }   -\tfrac{1}{4} g_{\mu \nu } F _{\rho \sigma }F ^{\rho\sigma  } .
\end{aligned}
\end{equation}
The equation of motion for the scalar field is 
\begin{equation}
    \nabla_{\mu} \nabla^{\mu} \phi=\frac{1}{4} \frac{d f(\phi)}{d \phi} F_{\mu \nu} F^{\mu \nu}~,
\end{equation}
while the equations for the Maxwell field are given by
\begin{equation}\label{eq:Maxwell}
    \nabla _{\mu } (f(\phi )F^{\mu \nu } )=0~.
\end{equation}

\section{\label{sec:level2}static solution}

In this section, we study the static equations of motion within a spherical symmetric spacetime directly. We use the   Painlevé-Gullstrand (PG) coordinates 
\begin{equation}
ds^{2} =-(1-\zeta ^{2} )\alpha ^{2}dt^{2}  +2\zeta \alpha dtdr+dr^{2}+r^{2}(d\theta ^{2}+sin^{2}\theta d\phi ^{2}). \label{eq:PG}
\end{equation} 
Here, metric functions $\zeta$ and $\alpha$ do not change over time and only depend on $r$. The apparent horizon $r_{h} $ is located where $\zeta (r_{h} )=1$. For a RN black hole, $\alpha =1$ and $\zeta =\sqrt{\frac{2M}{r}-\frac{Q^{2} }{r^{2} }  } $.

We take the gauge potential as $A_{\mu }dx^{\mu }=A(r)dt$. The Maxwell equations \eqref{eq:Maxwell} give
\begin{equation} \label{eq:Ar}
\partial _{r}A=\frac{Q\alpha }{r^{2}f(\phi ) },  
\end{equation}
in which $Q$ is the electric charge of the black hole. Then we get equations of motion for static background solutions
\begin{equation}
    \partial _{r} \alpha -r\alpha (\partial _{r}\phi  )^{2}=0 ,
\end{equation}
\begin{equation}\label{eq:jtzeta}
     \partial_{r} \zeta+\frac{\zeta}{2 r}-\frac{r\left(1-\zeta^{2}\right)}{2 \zeta}\left(\partial_{r} \phi\right)^{2}-\frac{Q^{2}}{2 r^{3} \zeta f(\phi)}=0,
\end{equation}
\begin{equation}\label{eq:jtphi}
 \begin{aligned}
    &\partial_{r}^{2} \phi+
    \frac{1}{\left(\zeta^{2}-1\right)}\Big[\left(\frac{Q^{2}}{r^{2} f(\phi)}+\zeta^{2}-2\right) \frac{\partial_{r} \phi}{r}-
\\
    &\frac{Q^{2}}{2 r^{4} f^{2}(\phi)} \frac{d f(\phi)}{d \phi}\Big]=0.
\end{aligned}  
\end{equation}

To solve these static equations, we   need to determine appropriate boundary conditions. At spatial infinity, the static solutions can be expressed as:
\begin{equation}\label{eq:wqy}
\begin{aligned}
    \zeta &=\sqrt{\frac{2M}{r} } \left ( 1-\frac{Q^{2}+Q_{s}^{2}   }{4Mr}+\cdots   \right ) ,\\
\phi &=\frac{Q_{s} }{r} +\frac{MQ_{s} }{r^{2} } +\cdots .
\end{aligned}
\end{equation}
Here $M$ is the total mass of the entire system, $ Q_{s}$ is  scalar charge. Near the event horizon $r_{h} $ of the black hole, the solutions can be expanded as:
\begin{equation}\label{eq:sjzeta}
    \zeta =1+\frac{1}{2r_{h} } \left ( \frac{Q^{2} }{f(\phi _{h} )r_{h}^{2}  } -1 \right ) (r-r_{h} )+\cdots ,
\end{equation}
\begin{equation}\label{eq:sjphi}
    \phi=\phi_{h}+\frac{Q^{2}}{2 r_{h} f\left(\phi_{h}\right)\left[Q^{2}-r_{h}^{2} f\left(\phi_{h}\right)\right]} \frac{d f\left(\phi_{h}\right)}{d \phi}\left(r-r_{h}\right)+\cdots,
\end{equation}
where $\phi_{h}$ is the scalar value on the event horizon. Given $\beta$, $Q$ and $M$, the scalar charge $Q_{s} $ or $\phi_{h}$ is determined.

By providing values for $\beta,Q,r_{h}$, and an initial guess value for $\phi_{h}$, and imposing the boundary conditions (\ref{eq:sjzeta} ,\ref{eq:sjphi}) at $r_{b1} =r_{h} (1+\epsilon )$, typically with $\epsilon\approx 10^{-7} $, we can numerically integrate the equations (\ref{eq:jtzeta},\ref{eq:jtphi}) up to $r_{b2}\approx 10^{7}r_{h}$. We need to adjust the initial guess value $\phi_{h}$ and repeat the integration process until the solution $\phi(r)$ approaches zero at $r_{b2}$. Upon securing the static background solutions $\zeta_{0}$ and $\phi_{0}$, we can employ equation \eqref{eq:wqy} to calculate the total mass $M$ and scalar charge $Q_{s}$, given by 
\begin{equation}
    M=\lim_{r \to \infty} \frac{r}{2} \zeta_0 (r)^{2} ,\quad Q_{s} =-\lim_{r \to \infty} r^{2} \partial _{r} \phi_0 (r). \label{eq:MQ}
\end{equation}

At spatial infinity, $\alpha$ can be expanded as:
\begin{equation}
    \alpha=a_{0}\left(1-\frac{Q_{s}^{2}}{2 r^{2}}+\cdots\right),
\end{equation}
where $a_{0}$  represents an undetermined constant due to the auxiliary freedom in $\alpha dt$ in PG coordinates. Here, we set $a _{0} =1$ to be consistent with the subsequent dynamic evolution method.

\begin{figure}[h]
\centering
\includegraphics[width=0.4\textwidth]{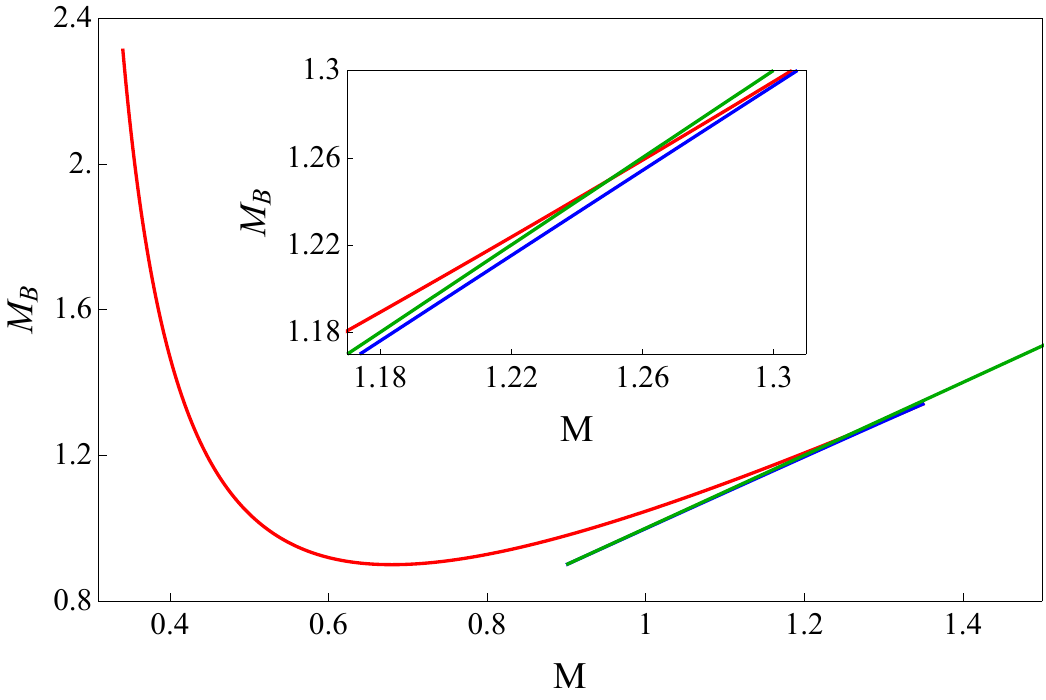}
\caption{The Christodoulou-Ruffini mass $M_{B}$ of the black hole is illustrated for static solutions with $\beta=200$ and $Q=0.9$. The horizontal axis represents the total mass $M$ of the system. The red and blue curves correspond to stable SBH and unstable SBH, respectively. The green curve represents the RN black hole. There are also three branches of static black hole solutions with other coupling parameter value, for example when $\beta=2000$, but the situation is similar to the case with $\beta=200$. However, $\beta$  can not be too small, otherwise there would be no SBH solutions in this model.}
    \label{fig:MB}
\end{figure}

In Fig. \ref{fig:MB}, we show the relation between the Christodoulou-Ruffini mass $M_{B}$ and the total mass $M$ of the static solutions.  In numerical relativity, one often uses the Arnowitt-Deser-Misner (ADM) mass which is defined at the spatial infinity as the total mass of the system. However, the ADM mass in PG coordinates always evaluates to zero and cannot reflect the correct physical mass of the spacetime \cite{shibata2015numerical}. Therefore, we employ the Misner-Sharp (MS) mass which is defined quasilocally in spherical symmetrical spacetimes as $m(t,r)=\frac{r}{2}\left(1-g^{\mu\nu}\partial_{\mu}r\partial_{\nu}r\right)$ where $r$ is the areal radius \cite{Misner:1964je,Hayward:1994bu}. It is easy to show that $m(t,r)=\frac{r}{2}\zeta^{2}$ in PG coordinates. At spatial infinity, one gets the total mass $M$ of the system as shown in equation (\ref{eq:MQ}). On the other hand, since $\zeta(t,r_{h})=1$ on the horizon, we have $m(t,r_{h})=\frac{1}{2}r_{h}$  which equals the irreducible mass of the black hole $M_{h}=\sqrt{\frac{S_{h}}{16\pi}}$ where $S_{h}=4\pi r_{h}^{2}$ denotes the area of the horizon. However, as the total mass $M$ keeps constant during the evolution while $M_h$ never decreases due to the area law, the masses defined above are not suitable for studying the energy extraction process.  In fact, in the study of superradiance, the usually adopted definition of black hole mass is the Christodoulou-Ruffini mass \cite{Zhang:2023qtn,East:2013mfa,Baake:2016oku,Okawa:2014nda,East:2017ovw,East:2018glu,Chesler:2018txn, Sanchis-Gual:2015lje,Sanchis-Gual:2016tcm,Bosch:2016vcp,Corelli:2021ikv}. For a spherical charged black hole, it is defined as $M_{B}=M_{h}+\frac{Q_{h}^{2}}{4M_{h}}=\frac{r_{h} }{2} +\frac{Q_{h}^{2} }{2r_{h}} $, where $Q_{h}$ is the charge within the horizon. In the model we considered here, there is $Q_h=Q$ since the scalar field is real and all the charge is contained in the horizon.

In Fig. \ref{fig:MB}, we select a RN black hole with total mass $M_0=1$ and charge $Q=0.9M_0$ as the reference black hole, and all other parameters are measured in units of $M_0$.
These solutions are obtained with fixed $Q = 0.9$ and $\beta=200$, and the total mass of the system varies. 
We have validated the correctness of the numerical static solutions by checking the Smarr relation \cite{Smarr:1972kt,Herdeiro:2018wub}
\begin{equation}
\label{eq:Smarr}
M^{2}+Q_{s}^{2}=Q^{2}+\frac{1}{4} S_{h}^{2} T_{h}^{2} .\end{equation}
Here $T_{h}=\frac{\alpha(r_h)}{2\pi}\zeta'(r_h)$ is the temperature of the black hole \cite{Jiang:2023yyn}. This relation is satisfied at the order $O(10^{-8})$.
As shown in Fig. \ref{fig:MB}, there are three different branches in the static solutions. The red branch corresponds to the linearly stable SBH, the blue branch represents the linearly unstable SBH, and the green branch is the linearly stable RN black hole. Note that RN black holes satisfy the bound $M>Q=0.9$, while  unstable SBH exists in the region $M\in(0.9,1.35)$, and more interestingly, stable SBH can have a total mass smaller than $Q$.

It is easy to show that $M_B=M$ for RN black holes. However, for SBHs, $M_B$ generally differs from $M$,  as shown in Fig. \ref{fig:MB}.  One might assume that $M$ should always be greater than $M_B$. However, this is not the case. The reason is rooted in the nonlocal property of the gravitational energy which extremely complicates the definition of gravitational energy. As a result, general relativity does not offer a single definition of the black hole mass, but offers several different definitions that are applicable under different circumstances. The total mass $M$ obtained from the equation (\ref{eq:MQ}) follows the definition of the Misner-Sharp mass, while the Christodoulou-Ruffini mass $M_{B}$ was introduced to study the energy extraction process from black holes. Since these masses are defined in different contexts, comparing their magnitudes is not so meaningful. From a practical standpoint, and to maintain consistency with prior research on superradiance \cite{Zhang:2023qtn,East:2013mfa,Baake:2016oku,Okawa:2014nda,East:2017ovw,East:2018glu,Chesler:2018txn, Sanchis-Gual:2015lje,Sanchis-Gual:2016tcm,Bosch:2016vcp,Corelli:2021ikv}, we use the Christodoulou-Ruffini mass $M_B$ to quantify the energy of the black hole. 
This definition was also used in \cite{Corelli:2021ikv} when investigating the superradiant process of RN black holes in the EMS model, while for SBHs they computed the mass at infinity from the stationary configuration that approximates the state of their simulation.

\begin{figure}[h]
\centering
\includegraphics[width=0.4\textwidth]{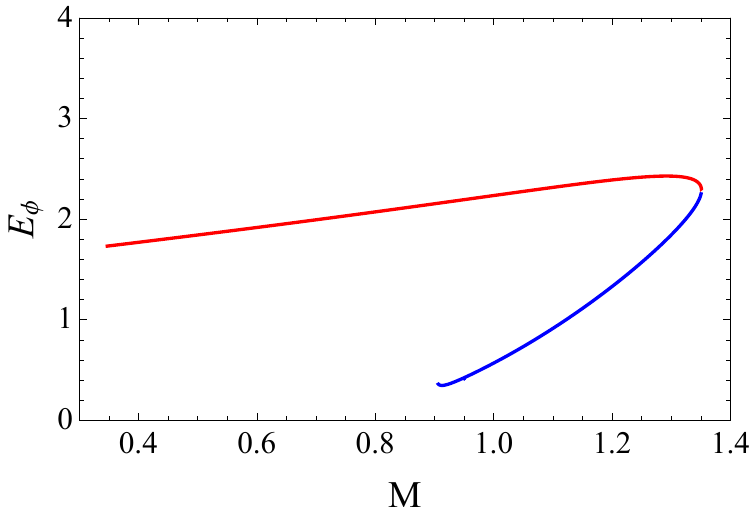}
\includegraphics[width=0.4\textwidth]{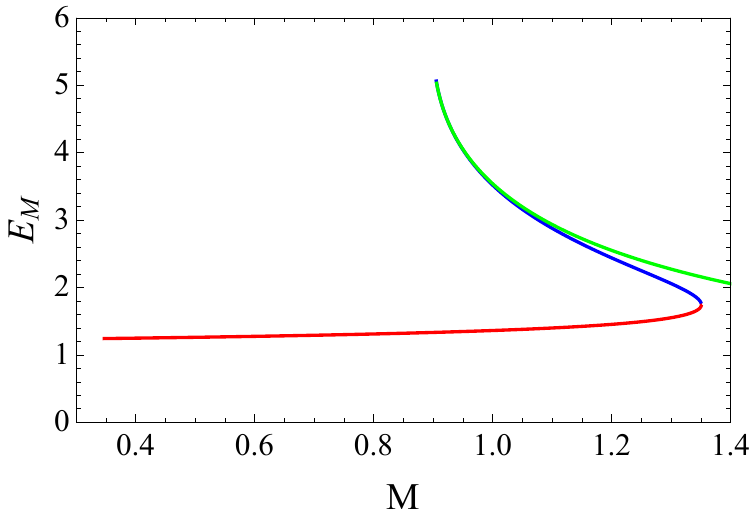}
\caption{The energy of the scalar field and Maxwell field outside the horizon. The parameters are the same as those in Fig. \ref{fig:MB}. The red branch represents the results for stable SBHs, the blue branch for unstable SBHs, and the green branch for the RN black holes. 
}
    \label{fig:Ephistatic}
\end{figure} 

On the other hand, to make the study more comprehensive, in Fig. \ref{fig:Ephistatic}, we show the energy of the scalar field and the Maxwell field of static solutions, which are defined as $
E_\phi=\frac{1}{4\pi}\int_{r_h}^\infty dV{T_{\mu\nu}^\phi} n^\mu n^\nu$ and $
E_M=\frac{1}{4\pi}\int_{r_h}^\infty dV{T_{\mu\nu}^A} n^\mu n^\nu$, respectively. Here $n_\mu=(-\alpha,0,0,0)$ is the normal vector of the constant time slice. 
The scalar field energy of both stable and unstable SBHs increases with increasing total mass. An RN black hole always has vanishing scalar field energy.  The Maxwell field energy of both unstable SBH and RN black holes decreases with $M$, while that of stable SBH increases with $M$. We would like to mention that $E_\phi$ and $E_M$ are also not defined in the same context as $M_B$. However, we could expect that if the black hole mass $M_B$ decreases, the scalar field energy $E_\phi$ or Maxwell field energy $E_M$ should increase. This is indeed observed in the following sections.



Now let us examine Fig. \ref{fig:MB} more carefully. In the coexistence region of three black hole solutions, the unstable SBH always has the smallest $M_B$. On the other hand, stable SBH has the largest $M_B$ when $M<1.25$, while RN BH has the largest $M_B$ when $M>1.25$. These observations suggest that in principle, if one can transfer the stable SBH to the RN BH (or the opposite process) in an appropriate mass region through dynamical first-order phase transition, the black hole mass $M_B$ can be decreased. Moreover,  stable SBH can possess a very large $M_B$ when the total mass $M$ is small (for example when $M<0.6$). This implies that if we applied perturbation to the stable SBH, as the total mass $M$ increases with the perturbation, the black hole mass $M_{B}$ would decrease. In the following, we will numerically simulate the dynamical evolution of black holes to verify these conjectures.



\section{\label{sec:level3} NUMERICAL SETUP FOR DYNAMICS}

In this section, we focus on the full nonlinear dynamical evolution of black holes under perturbation in a spherically symmetric spacetime. 
In the PG coordinates (\ref{eq:PG}), the metric functions $\zeta$ and $\alpha$ now vary over time and radial radius $(t,r)$. 
Introducing two auxiliary variables 
\begin{align}
\Phi &=\partial _{r}\phi, \label{eq:Phi}
\\
\Pi &=\frac{1}{\alpha }\partial _{t}\phi -\zeta \Phi. \label{eq:Pi} 
\end{align}
The Einstein equations give
\begin{equation}\label{eq:rzeta}
\partial _{r}\zeta =\frac{r}{2\zeta } (\Phi ^{2}+\Pi ^{2}  )+\frac{Q^{2} }{2r^{3}\zeta f(\phi ) }+r\Pi \Phi -\frac{\zeta }{2r},
\end{equation}
\begin{equation}\label{eq:ralpha}
\partial _{r}\alpha =-\frac{r\Pi \Phi \alpha }{\zeta },  
\end{equation}
\begin{equation} 
\label{eq:tzeta}
\partial _{t} \zeta =\frac{r\alpha }{\zeta } (\Pi +\Phi \zeta )(\Pi \zeta +\Phi ).
\end{equation}
The scalar equations can be written as:
\begin{equation}\label{eq:tphi}
\partial _{t}\phi =\alpha (\Pi +\Phi \zeta ) ,
\end{equation}
\begin{equation}
\label{eq:tPi}
\partial _{t} \Pi =\frac{\partial _{r}[(\Pi\zeta +\Phi )\alpha r^{2} ] }{r^{2} }+\frac{\alpha }{2}\frac{Q^{2} }{r^{4}f^{2}(\phi )  }\frac{df(\phi )}{d\phi } ,  \end{equation}

We use these equations to solve the metric functions $\zeta,\alpha$ and the scalar field functions $ \phi,\Phi, \Pi$. If the initial scalar distribution $\phi$ and $\Pi$ are given, we can get the initial $\Phi$ and $\zeta,\alpha$ from Eq.(\ref{eq:Phi},\ref{eq:rzeta},\ref{eq:ralpha}). The $\zeta,\phi,\Pi$ on next time slices can be obtained from the evolution Eq.(\ref{eq:tzeta},\ref{eq:tphi},\ref{eq:tPi}), respectively. 
The values of $\Phi$ and $\alpha$ can be derived by applying constraint Eq.(\ref{eq:Phi},\ref{eq:ralpha}), respectively. By repeating this iterative procedure, we can determine all the metric and scalar functions at each time step. 
The constraint given by Eq.(\ref{eq:rzeta}) is applied solely at the outset.

At spatial infinity, the matter field functions $\phi,\Pi,\Phi$ should be zero. 
Then $\left.\zeta \right|_{r \rightarrow \infty}=\sqrt{2M/r } $, in which the constant $M$ is the total Misner-Sharp mass \cite{Misner:1964je} of the spacetime. However, we encounter challenges of $\zeta$ decay. To ensure the stability and long-term evolution of the numerical simulation, we introduce $s=\sqrt{r}\zeta  $ to replace $\zeta$ in the simulation, and set the boundary conditions for $s,\alpha$ as 
\begin{equation}
s\mid _{r\to \infty }=\sqrt{2 M}, ~\alpha \mid _{r\to \infty }=1.\end{equation}
The computational domain of the system ranges within $[r_{in},\infty  )$, where $r_{in}$ is slightly less than the initial apparent horizon $r_{h}$. 
Moreover, we employ the fourth-order finite difference method in the radial direction through compactification $z=\frac{r}{1+r} $ \cite{Ripley:2019aqj} and discretizing $z$ uniformly with approximately $2^{11} \sim 2^{12} $ grid points. 
So the system is evolved within the region $z\in [z_{in},1)$, where $z = 1$ corresponds to the spatial infinity. The time evolution is solved with the fourth-order Runge-Kutta method. 
To ensure stability in the simulation, we employ Kreiss-Oliger dissipation.
In the first step, the constraint \eqref{eq:rzeta} is solved using the Newton-Raphson method \cite{Dias:2015nua}. 

If seed black holes are stable and unstable SBHs, we take ingoing initial scalar perturbation with the following form 
\begin{equation}
\begin{aligned}
\delta \phi(r) &=p\begin{array}{l} 
  \left\{\begin{matrix} 
   e^{-\frac{1}{r-r_{1} }-\frac{1}{r_{2}-r}  }(r-r_{1} )(r_{2}-r ) ,r_{1}<r<r_{2},      \\ 
 0,  \quad\quad\quad\quad\quad\quad\quad\quad\quad\quad\quad\quad\text{otherwise},
\end{matrix}\right.    
\end{array} 
\\
\Pi (r)&=\partial _{r}\delta \phi (r)-\zeta _{0}(r)\partial _{r}\phi _{0} (r) .
\end{aligned} \label{eq:deltaphi1}
\end{equation}
in which $r_{1}=3,r_{2}=9  $ and $p$ is the perturbation amplitude parameter. 

If the seed black hole is a RN black hole, 
the ingoing initial scalar perturbation gives
\begin{equation}
\begin{aligned}
\delta \phi(r) &=p\begin{array}{l} 
  \left\{\begin{matrix} 
   10^{-2} e^{-\frac{1}{r-r_{1} }-\frac{1}{r_{2}-r}  }(r-r_{1} )^{2}(r_{2}-r )^{2} ,r_{1}<r<r_{2}      \\ 
 0,  \quad\quad\quad\quad\quad\quad\quad\quad\quad\quad\quad\quad\text{otherwise},
\end{matrix}\right.    
\end{array} 
\\
\Pi (r)&=\partial _{r}\delta \phi (r).
\end{aligned} \label{eq:deltaphi2}
\end{equation}
in which $r_{1}=4,r_{2}=9  $.

\section{NUMERICAL RESULTS}

\subsection{The initial black hole is a RN black hole}

\begin{figure}[h]
\centering
\includegraphics[width=0.4\textwidth]{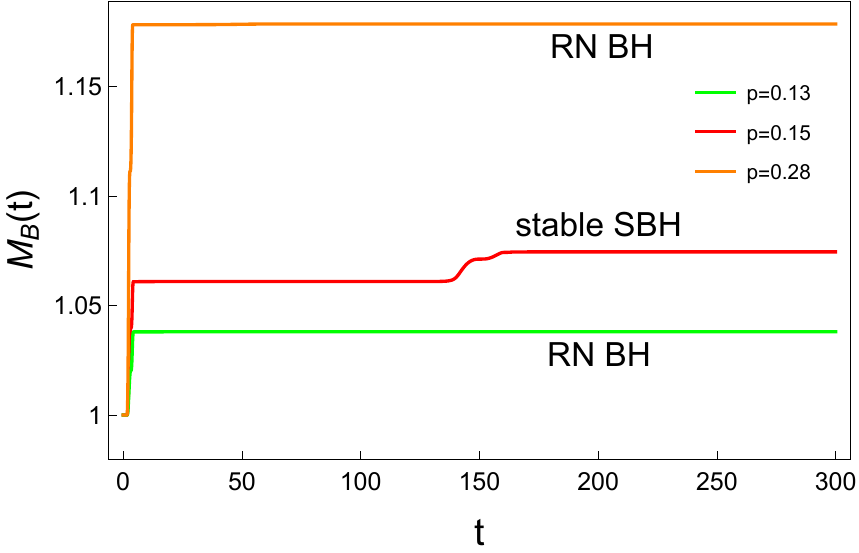}
\caption{The evolution of black hole mass $M_{B}$ starting from an RN BH with $\beta =2000,M=1,Q=0.9,M_{B}=1$ under perturbation (\ref{eq:deltaphi2}). The black hole mass always increases here.
}
    \label{fig:b2000RNCR}
\end{figure}

\begin{figure}[h]
\centering
\includegraphics[width=0.4\textwidth]{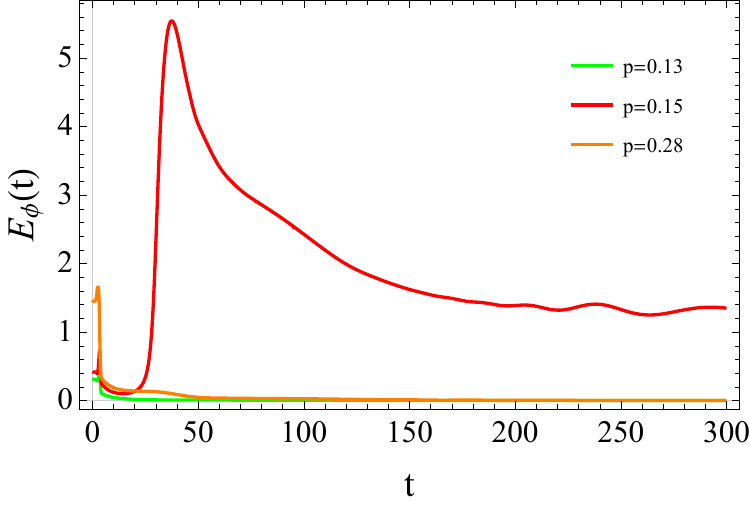}
\includegraphics[width=0.4\textwidth]{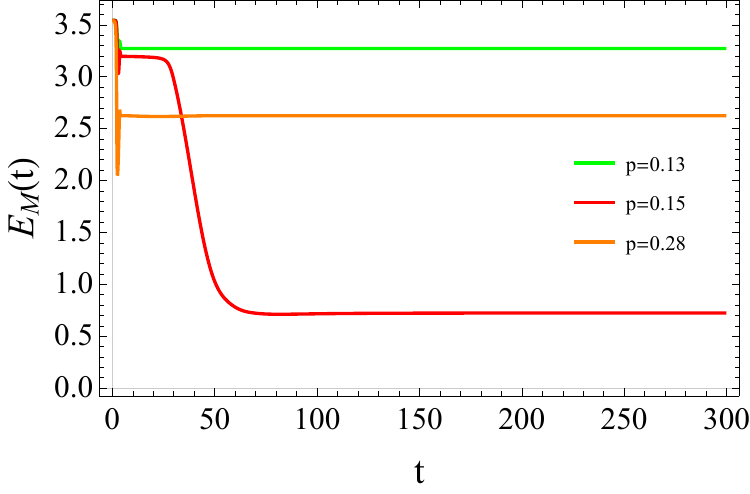}
\caption{The evolution of scalar field energy $E_{\phi }$ and Maxwell field energy $E_M$ starting from an RN BH with $\beta =2000,M=1,Q=0.9,M_{B}=1$ under perturbation (\ref{eq:deltaphi2}). Note that the later fluctuations in $E_\phi$ when $p=0.15$ are caused by numerical errors in the integration (\ref{eq:EphiMt}). 
}
    \label{fig:b2000RNEphi}
\end{figure}

Let us start with the RN black hole  with $M=1,Q=0.9$ under  perturbation (\ref{eq:deltaphi2}). 
To make it easier to observe the dynamical transition to SBH, we take the coupling with $\beta=2000$ in this subsection.  
The dynamic results are shown in Fig. \ref{fig:b2000RNCR} and Fig. \ref{fig:b2000RNEphi}. 
When the perturbation is not large enough, the RN black hole maintains its stability, as depicted by the green curve.
However, as the perturbation intensifies, the stability of the RN black hole is destroyed. 
Consequently, the RN black hole transitions into a SBH, as shown by the red curve. 
Yet, when the perturbation becomes excessively large, the ultimate state reverts to a RN black hole. 
This outcome aligns with the static solutions presented in Fig. \ref{fig:MB}, in which only RN solution exists for very large mass $M$.


Let us now observe the evolution of the black hole mass. 
In our simulations, we find that after perturbing the seed black hole, the black hole mass $M_{B}$ suddenly rises and then saturates. The initial surge  is attributed to the absorption of the scalar perturbation. The later saturation implies that the system has reached a new equilibrium state.  The greater the scalar perturbation, the greater the black hole mass of the final solution. 
We wish to emphasize that, following our simulations of the evolution of RN black holes with a variety of parameters, we have not observed any energy extraction. This holds true regardless of whether the final state of the system remains a RN black hole or transitions to a SBH.

Fig. \ref{fig:b2000RNEphi} shows the corresponding evolution of the scalar field energy $E_{\phi}$ and Maxwell field energy $E_M$, which have the following form in the dynamical case:
\begin{equation}
E_{\phi}=\frac{1}{2}\int_{r_h}^\infty dr (\Pi^2+\Phi^2), E_M=\frac{1}{2}\int_{r_h}^\infty dr \frac{Q^2}{r^2f(\phi)^2}. \label{eq:EphiMt}
\end{equation}
If the final solution is still a RN black hole, $E_{\phi}$ decreases to zero at late times because the scalar field is absorbed by the black hole, and $E_M$ becomes smaller than the initial value as $r_h$ increases in the integration (\ref{eq:EphiMt}). On the other hand, if the final solution transitions to a SBH, $E_{\phi}$ gets a nonvanishing final value, meanwhile $E_M$ is significantly reduced as $f(\phi)$ modulates the energy of Maxwell field, as shown in \eqref{eq:EphiMt}. This indicates that during the scalarization process, some energy of the Maxwell field is converted into the scalar field, and some is absorbed by the black hole.

\subsection{The initial black hole is an unstable SBH}

In previous work \cite{Jiang:2023yyn}, RN black holes and stable SBH have been employed as initial configurations to study the critical dynamics of first-order phase transitions in black holes. 
However, the use of an unstable SBH as an initial configuration for dynamic evolution has not yet been explored. 
Here, we take the unstable SBH with $M=1,Q=0.9,M_B=0.99$ as the initial configuration, which lives in the blue branch in Fig. \ref{fig:MB}.  We apply an ingoing scalar field perturbation  \eqref{eq:deltaphi1} to the systems.  Since this seed black hole is linearly unstable, the dynamical transition can occur under arbitrarily small perturbation, which can drive it to one of two possible stable final states: either the stable SBH on the red branch or the RN black hole on the green branch in Fig. \ref{fig:MB}.

\begin{figure}[h]
\centering
\includegraphics[width=0.4\textwidth]{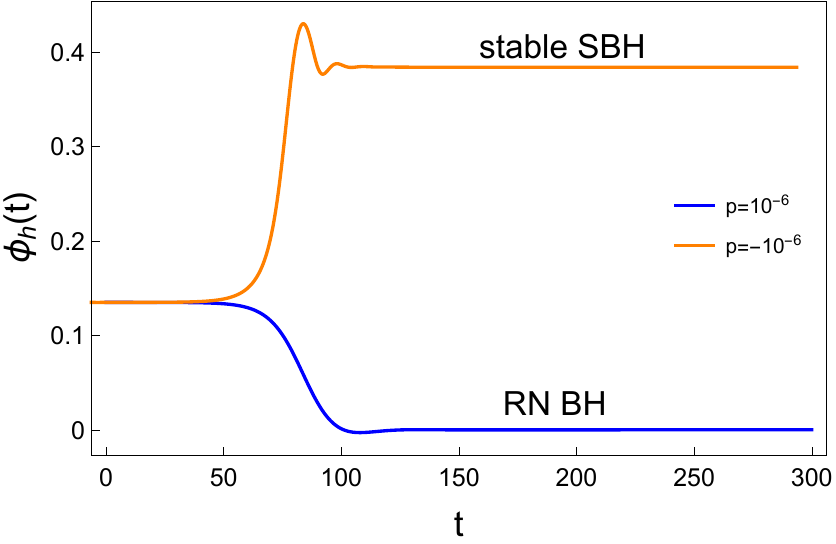}
\caption{The value of the scalar field on the horizon $\phi _{h} (t)$ as a function of time in the case where the perturbation amplitude $p$ is close to zero  for initial data family \eqref{eq:deltaphi1}. 
We take the unstable SBH with $M=1,Q=0.9,M_B=0.99$ as the initial configuration.}
    \label{fig:phihp0}
\end{figure}

In Fig. \ref{fig:phihp0}, we show the evolution of the value of the scalar field on the apparent horizon $\phi_h(t)$. It can be seen that the perturbation amplitudes $p$ with different signs can induce the system to evolve into two different stable final states. The absolute value of the amplitude $p$ here is  approximately equal to $10^{-6} $, corresponding to a very small perturbation. The final state of the dynamical process corresponding to the positive perturbation amplitude ($p>0$) is an RN black hole. In contrast, the negative perturbation amplitude ($p<0$) pushes the evolved system to a stable SBH. This result indicates that the gravitational system undergoes a special class of critical dynamics with a perturbation strength threshold of zero $p_{0}=0$. In this case, the critical state in the critical dynamical process is the initial state itself. The smaller the perturbation strength, the longer the system will remain in the unstable initial state.

\begin{figure}[h]
\centering
\includegraphics[width=0.4\textwidth]{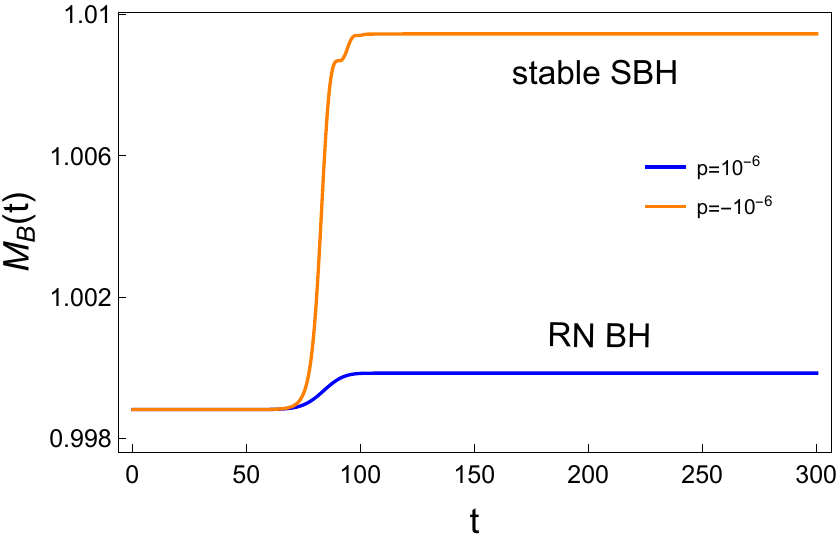}
\caption{The evolution of Christodoulou-Ruffini mass $M_{B}$ after the perturbation \eqref{eq:deltaphi1} with 
$p=\pm 10^{-6} $ of the unstable SBH. The black hole mass always increases here.}
    \label{fig:coldMB0}
\end{figure}

\begin{figure}[h]
\centering
\includegraphics[width=0.4\textwidth]{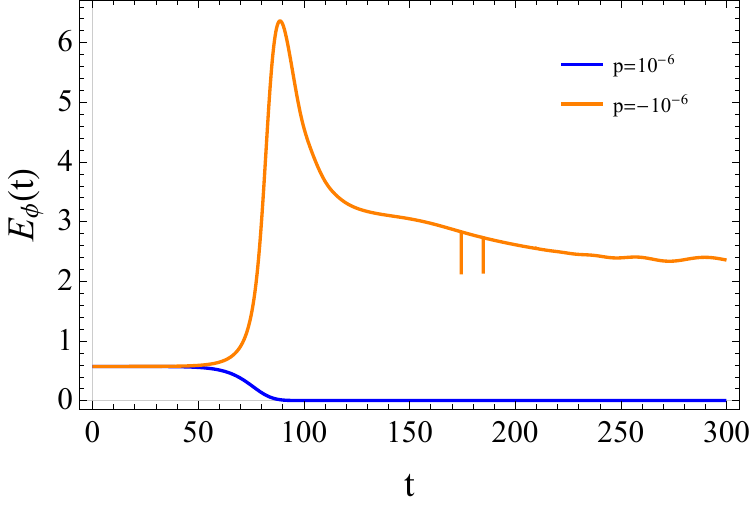}
\includegraphics[width=0.4\textwidth]{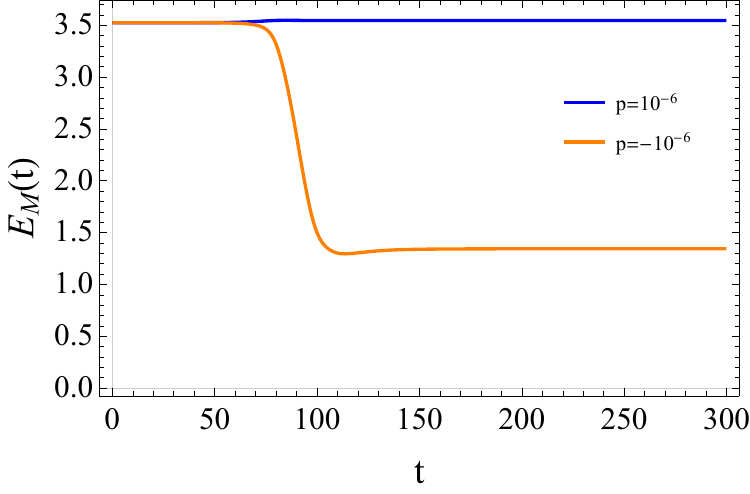}
\caption{The evolution of scalar field energy $E_{\phi }$ and Maxwell field energy $E_M$ after the perturbation \eqref{eq:deltaphi1} with 
$p=\pm 10^{-6} $ of the unstable SBH. The later fluctuations in $E_\phi$ when $p=-10^{-6}$ are caused by numerical errors in the integration (\ref{eq:EphiMt}).}
    \label{fig:cold0Ephi}
\end{figure}

From Figs. \ref{fig:coldMB0} and \ref{fig:cold0Ephi}, regardless of whether the scalar perturbation is positive or negative, the black hole mass $M_{B}$ always increases. When the unstable SBH evolves to a RN black hole, part of the scalar field energy is absorbed by the black hole, while the other part is transferred to the Maxwell field.  
When the unstable SBH evolves to a stable SBH, some of the Maxwell field energy is transferred to the scalar field, while some is absorbed by the black hole.
In both cases, the final black holes have a larger mass $M_B$ than the seed unstable SBH. 
This confirms the observation for the unstable SBH in Fig. \ref{fig:MB}.

 \begin{figure}[h]
\centering
\includegraphics[width=0.4\textwidth]{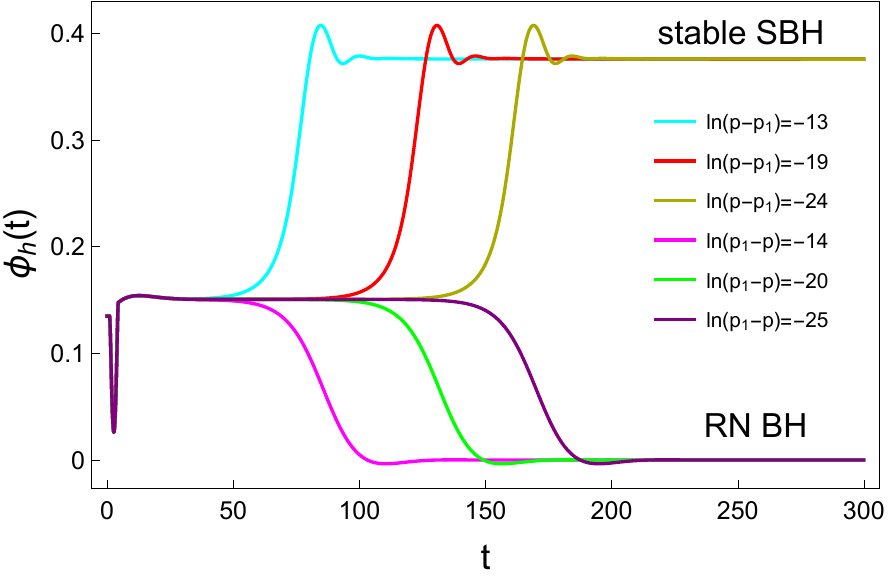}
\caption{The evolution of the scalar field value $\phi_{h}$ on the apparent horizon with respect to various $p$ near threshold $p_{1} \approx  -0.00631188663791$ for initial data family \eqref{eq:deltaphi1}. We take the unstable SBH with $M=1,Q=0.9,M_B=0.99$ as the seed black hole.}
    \label{fig:phihp1}
\end{figure}

Interestingly, as the perturbation amplitude $p$ gradually decreases, we find that besides $p_{0}=0$, there is another critical value $p_{1}\in (-0.0064,-0.0063)$ for the perturbation amplitude, which divides the final state of evolution into two parts, as shown in Fig. \ref{fig:phihp1}.
Through the binary method, we continuously approach this critical value $p_{1}\approx -0.00631188663791 $. As $p$ approaches $p_{1}$ from either below or above, after experiencing a rapid change in the early stages of evolution, all the intermediate solutions are attracted to a plateau. 
The closer $p$ is to $p_{1}$, the longer $\phi_{h}$ remains on this plateau. Essentially, the plateau signifies a linearly unstable static SBH. 
When the negative perturbation amplitude $p_{1}<p<0 $, the evolutionary final state is a stable   SBH, while if $p< p_{1} $, the final black hole undergoes descalarization and becomes a RN black hole.

\begin{figure}[h]
\centering
\includegraphics[width=0.4\textwidth]{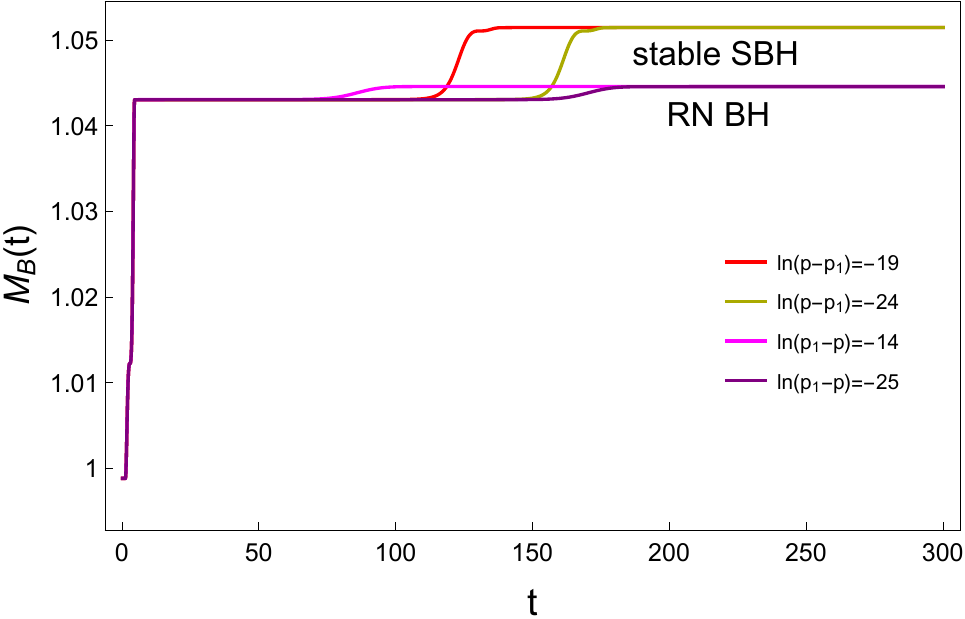}
\includegraphics[width=0.4\textwidth]{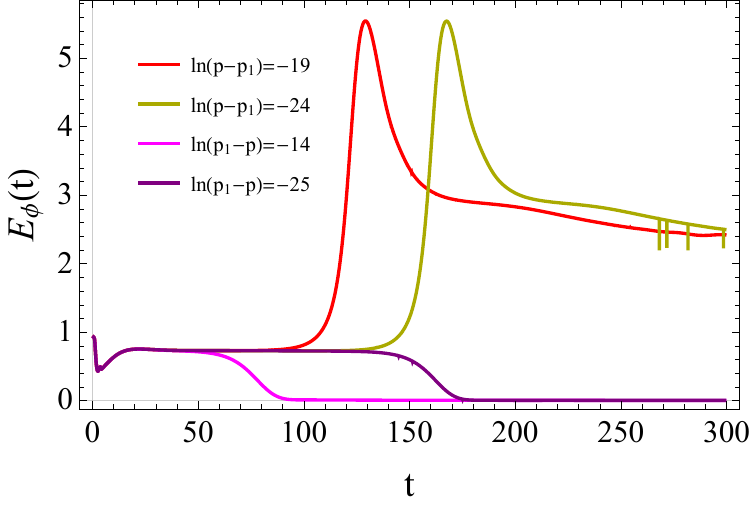}
\includegraphics[width=0.4\textwidth]{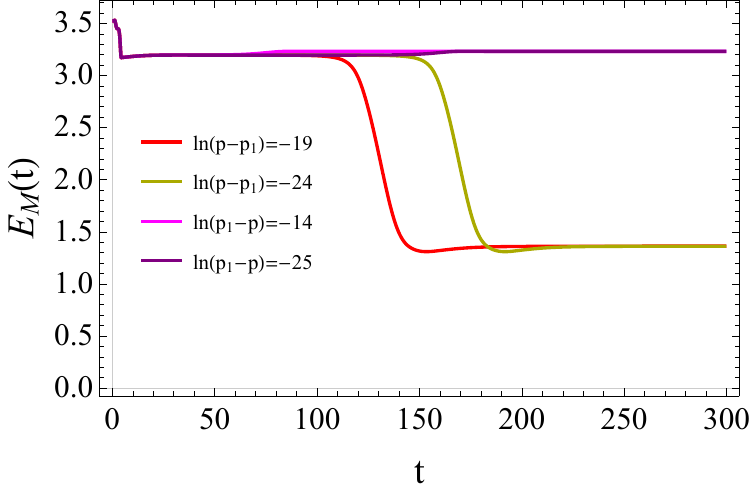}
\caption{The evolution of black hole mass $M_{B}$, scalar field energy $E_{\phi }$ and Maxwell field energy $E_M$ with respect to various $p$ near threshold $p_{1} \approx  -0.00631188663791$ for initial data family \eqref{eq:deltaphi1}. The seed black hole is an unstable SBH with $M=1,Q=0.9,M_B=0.99$. The later fluctuations in $E_\phi$ are caused by numerical errors in the integration (\ref{eq:EphiMt}).}
    \label{fig:coldcriEphi}
\end{figure} 

The plateau is also observed in Fig. \ref{fig:coldcriEphi}. The duration of $M_{B}$ staying on this plateau becomes longer as $p$ approaches $p_{1}$. 
Ultimately, the evolved black holes, whether they are stable SBH or RN black holes, possess a greater mass $M_{B}$ compared to the initial unstable SBH.
The behaviors of $E_\phi$ and $E_M$ are similar to those shown in Fig. \ref{fig:cold0Ephi}. When the black hole evolves into a RN black hole, part of the scalar field energy is  absorbed by the black hole, while the remaining part is transferred to the Maxwell field. 
When the black hole evolves into a stable SBH, the Maxwell field energy is significantly reduced. Some are transferred to the scalar field, and some are absorbed by the black hole.

\subsection{The initial black hole is a stable SBH} \label{sec:SBHMB}

\begin{figure}[h]
\centering
\includegraphics[width=0.4\textwidth]{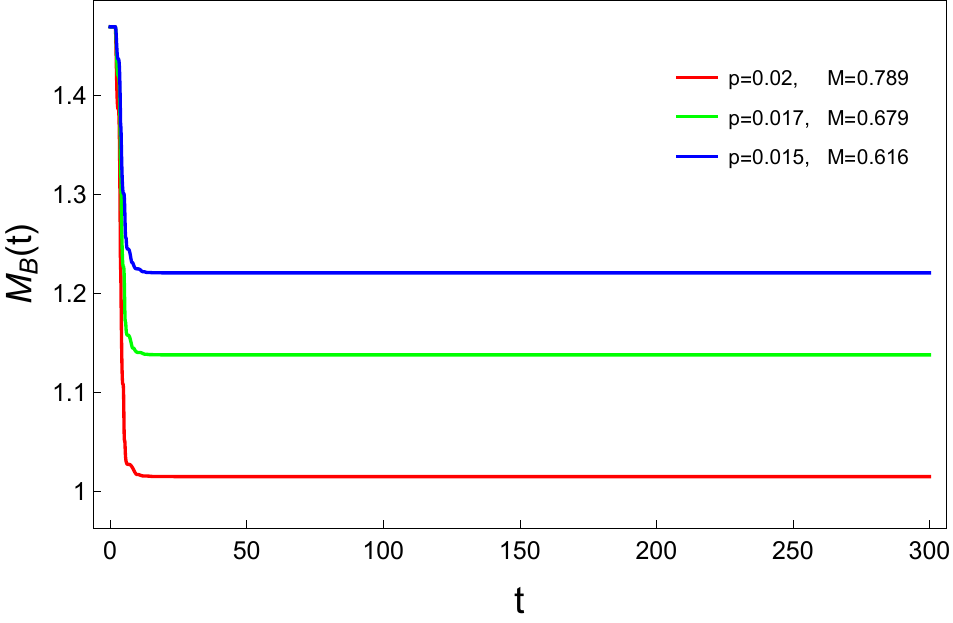}
\caption{The evolution of black hole mass $M_B$ when the seed black hole is a stable SBH with $M=0.4,Q=0.9,M_B=1.46$ under the perturbation \eqref{eq:deltaphi1} with various $p$. The black hole mass all decreases here.}
    \label{fig:hotCR}
\end{figure}

Now we take a small stable SBH with $M=0.4,Q=0.9,M_B=1.46$ as the seed black hole  under perturbation (\ref{eq:deltaphi1}). 
As depicted in Fig. \ref{fig:hotCR},
the mass $M_B$ of the final  black hole  decreases with increasing perturbation. 
This implies that a portion of the energy of the black hole is extracted. Since the scalar field is real, the black hole charge cannot be extracted out of the black hole. Then according to the formula $M_B=\frac{r_h}{2}+\frac{Q^2}{2r_h}$, the decrease in $M_B$ is mainly caused by the variation in the horizon radius $r_h$. Actually,  the black hole mass has a minimal value $M_B= Q$ at $r_h=Q$. When $r_h<Q$, the black hole mass $M_B$ decreases with $r_h$.  Since our small seed SBH has $r_h\le Q$, thus $M_B$ decreases with increasing perturbation.  This result resembles the evolution of scalar field around the black hole in  Einstein-scalar-Gauss-Bonnet gravity  \cite{East:2020hgw}.  

\begin{figure}[h]
\centering
\includegraphics[width=0.4\textwidth]{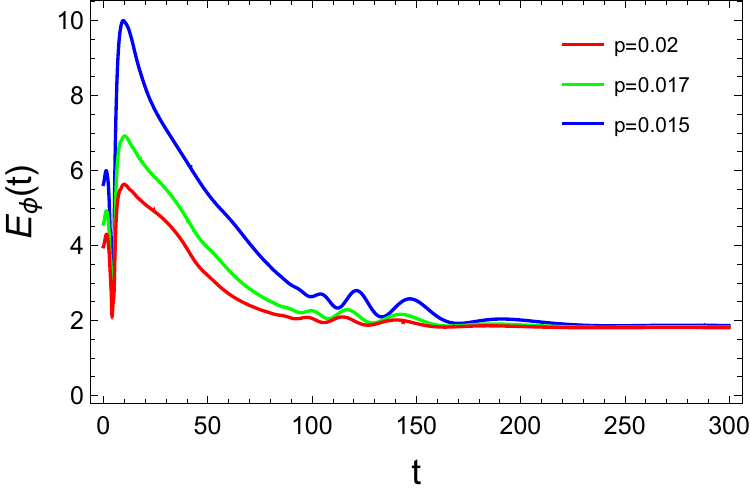}
\includegraphics[width=0.4\textwidth]{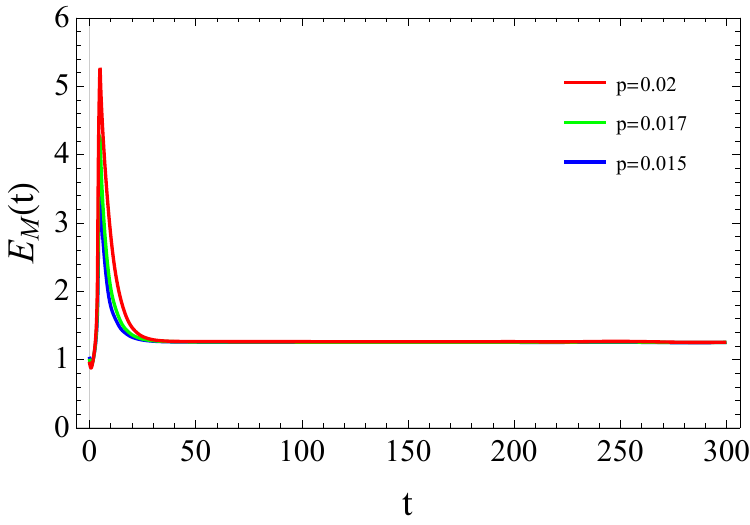}
\caption{
The evolution of scalar field energy $E_{\phi }$ and Maxwell field energy $E_M$ when the seed black hole is a stable SBH with $M=0.4,Q=0.9,M_B=1.46$ under the perturbation \eqref{eq:deltaphi1} with various $p$.}
    \label{fig:hotEphi}
\end{figure}

In Fig. \ref{fig:hotEphi}, we show the corresponding evolution of the scalar field and Maxwell field energy. The scalar field energy decreases, while the Maxwell field energy increases. In some sense, this indicates that the increased Maxwell field energy comes from the scalar field and the central black hole. But this statement is only qualitative, not quantitative, as $M_B$ are not defined in the same context as $E_\phi$ and $E_M$. 

On the other hand, as we have mentioned in section \ref{sec:level2}, there is another possible way through the first-order phase transition to extract the black hole energy when the stable seed  SBH has a relatively large $M$. However, this situation is not found in the dynamical evolution. Since the  stable SBH  and RN black holes are both stable, only a large perturbation can trigger the transition between them.  However, a large perturbation leads to a significant increase in total mass $M$ of the system, and the $M_B$ of the corresponding final black hole always has a larger value than the seed black hole.

\section{CONCLUSION}

We have studied the evolution of black hole energy in EMS theory. 
From the static solution analysis, it can be seen that adding perturbations to the RN black hole may reduce the black hole mass, but this situation has not been found in the dynamic evolution. 
We  have also taken the seed black hole as an unstable SBH, but found the  black hole mass always increases. 
In this case, we find that in addition to the critical perturbation amplitude $p_{0}=0$, there exists another critical amplitude $p_{1}<0$. 
If perturbation amplitude deviates slightly from the critical value,  the black hole  eventually evolves into a stable SBH or a RN black hole.

We have investigated the dynamical evolution of a stable SBH. 
When the initial total mass is relatively small, applying perturbation to it extracts the black hole energy, resulting in a decrease in black hole mass.  This is consistent of the analysis from the static solutions.
When the total mass of the stable seed SBH is relatively large, the black hole undergoes a first-order phase transition  and eventually transitions to a RN black hole. The black hole mass may decrease, but we do not find this situation in numerical simulations. This implies that the dynamical first-order phase transition is not always
accompanied by black hole energy extraction.

We should mention that the decrease in black hole energy in the EMS model is not related to superradiance. Here the scalar field extracts black hole energy that resembles those found in the Einstein-scalar-Gauss-Bonnet  model \cite{East:2020hgw}.
In addition, there may also be phenomena of black hole energy extraction in other models with first-order phase transitions \cite{Liu:2022fxy}.  


\begin{acknowledgments}
This work is supported by the Natural Science Foundation of China (NSFC) under Grant No.  12375048.
\end{acknowledgments}

\appendix

\nocite{*}

\bibliography{writing}

\end{document}